\documentclass[a4paper,12pt]{article}
\begin{document}
\newcommand{\eq}[1]{~(\ref{#1})}
\newcommand{\BEQ}{\begin{equation}}
\newcommand{\EEQ}{\end{equation}}
\newcommand{\norm}[1]{\mbox{$\left| \left| #1 \right| \right| $}}
\newcommand{\abs}[1]{\mbox{$\left| #1 \right| $}} 
\newcommand{\bra}[1]{\mbox{$\langle \left. #1 \right| $}}
 \newcommand{\ket}[1]{\mbox{$\left|  #1 \rangle \right. $}}
\newcommand{\DEG}[1]{\mbox{$ #1^{\rm o}$}}
\newcommand{\lappr}{\mbox{$\stackrel{<}{\sim}$}} 
\newcommand{\gappr}{\mbox{$\stackrel{>}{\sim}$}} 
\newcommand{\mr}[1]{\mbox{\rm #1}} 
\newcommand{\ETC}{\mbox{\em etc.\/ }}
\newcommand{\VIZ}{\mbox{\em viz.\/ }}
\newcommand{\CF}{\mbox{\em cf.\/ }}
\newcommand{\IE}{\mbox{\em i.e. \/}}
\newcommand{\ETAL}{\mbox{\em et. al.\/ }}
\newcommand{\EG}{\mbox{\em e.g.\/ }}
%
%
\begin{flushright}
DFF-280/04/1997 (Florence)\\
JHU--TIPAC--97007 (Johns Hopkins)\\
April  1997 
\end{flushright}
\vspace*{8mm}
\begin{center}
{\Large\bf Tests of Basic Quantum Mechanics in Oscillation
Experiments.}\\[5truemm]
G. Domokos and S. Kovesi--Domokos\\[2mm]
Dipartimento di Fisica, Universit\'{a} di Firenze\\
Florence, Italy\\[1mm]
and\\[1mm]
The Henry A. Rowland Department of Physics and Astronomy\\
The Johns Hopkins University\footnote{Permanent address. 
E--mail:~SKD@HAAR.PHA.JHU.EDU}\\
Baltimore, MD 21218
\end{center}
\vspace*{4mm}
\begin{quote}
According to standard quantum theory, the time evolution operator
of a quantum system is independent of the state of the system. One
can, however, consider systems in which this is not the case: the
evolution operator may depend on the density operator itself. 
The presence of such modifications of quantum theory can be tested
in long baseline oscillation experiments.
\end{quote}
\vspace*{5mm}

While quantum theory works remarkably well and there is no evidence to
date that its validity may be limited, there have been attempts at
modifying its structure both in order to resolve some conceptual
problems or in order to establish its limits of validity. 
In particular, modifications of quantum theory have been 
considered in order to resolve the problem of transition to
classical theory (the problem of ``decoherence''). Briefly,
it has to be made sure that if two states are macroscopically
different, then there is no definite phase relation between them
and thus, their superposition is not a physically admissible
state. (Sometimes this is dubbed as the ``problem of Schr\"{o}dinger's
cat'', see for instance D'Espagnat~\cite{despagnat}.) It is generally
thought that decoherence is a consequence of some kind of an
interaction with the environment in some general sense. 
For instance, it was conjectured that gravitational
interactions cause decoherence once a body
is sufficiently massive~\cite{karolyhazy}.   However, it is also conceivable
that decoherence is an intrinsic property
of quantum theory itself: after a sufficiently long lapse of
time, phase relations are lost even between microscopic 
states and, therefore, macroscopic ones as well ({\em
``spontaneous decoherence''}). Clearly, this
requires a modification of the formalism of quantum theory as we 
know it.
Such modifications have been repeatedly advocated by a variety of
authors; a reasonably complete and  up-to-date account of the
problem of decoherence is given in the book by Omn\`{e}s~\cite{omnes}.

In this note
we consider one class of  such possible modifications and suggest that one
can place limits on departures from the standard version of quantum
theory in experiments currently being carried out or being planned.

The standard form of the time evolution operator of a quantum system
described by its density operator, $\rho$ is given by the expression
\footnote{Natural units are used: $\hbar = c =1$}:
\BEQ	
i\frac{\partial \rho}{\partial t} = \left[ H , \rho \right],
\label{schrodinger}
\EEQ
$H$ being the Hamiltonian of the system. Equation \eq{schrodinger}  
is linear in $\rho$. This fact has 
remarkable consequences; perhaps the most important one is that
if at a time a system was in a pure state (\IE
$\rho^{2}\left( t_{0}\right)= \rho \left( t_{0}\right)$), then it will 
remain in a pure state at all times. (Conversely, a statistical
mixture will not evolve  spontaneously into a pure state either.)

It is conceivable, however that, due to the presence of some, hitherto
undetected small non linearity in equation \eq{schrodinger}, the  
property we just described is only an approximate one: if left alone
for a sufficiently long  time, the system spontaneously evolves 
into a mixture or a mixture contracts to some pure state.

In order to explore this possibility, we add a non linear term
to the right hand side of the evolution equation, \eq{schrodinger}
and from now on investigate the evolution of a density operator
 obeying the equation:
\BEQ
i\frac{\partial \rho}{\partial t} = \left[ H, \rho \right]
 -i\frac{a}{T} \left( f(\rho) - \frac{1}{N}{\rm Tr} f(\rho)\right)
\label{something}
\EEQ
Here $T$ is some characteristic time scale. The 
constant $a$ serves the purpose of normalizing the
time dependence of the solutions due to the presence
of the term added to \eq{schrodinger}. The function $f(\rho)$ 
 governs the deviation of the evolution from standard
quantum mechanics. (Not having any firm guiding principle,
we shall experiment with some simple functional forms.)
 Here and for the rest of this paper we
restrict ourselves to state spaces of dimension $N$. A 
generalization to infinite dimensional
state spaces appears to be straight forward; however, we have not
explored it in detail.

In writing down \eq{something} we were guided by some physical
prejudices. Three of those are worth noting.
\begin{itemize}
\item We want to maintain probability conservation, therefore
the term added to the right hand side of \eq{schrodinger} is
traceless. As a consequence, 
\[ \frac{\partial {\rm Tr}\rho}{\partial t}=0, \]
as in standard quantum mechanics.
\item We wrote down the  evolution equation in such a manner that it
contains a characteristic time scale governing the deviation from
standard quantum mechanics. (Alternatively, we could have, \EG
contemplated a deformation of the Heisenberg algebra. However,
deformations of Lie algebras as discussed in the literature, contain
dimensionless parameters. We consider $T$ a constant of Nature; there
are no dimensionless constants of Nature we know of.\footnote{Often,
the fine structure constant is considered a constant of Nature. It is
not, however, because its magnitude depends on the energy scale at
which the measurement is performed.})
\item The term added to the evolution equation is local in time.
One could have given, \EG some memory to the system by making the
evolution equation depend on $\rho$ taken at some past moment
or upon an integral of $\rho$, \ETC However, if eventually, one
wants to construct manifestly Lorentz invariant theories, such terms 
are hard or impossible to incorporate. 
\end{itemize}

We note that the class of modifications considered here is, in a sense,
{\em a minimal one\/:} apart from the linearity of the evolution
equation, no other essential property of quantum theory (\EG the
superposition principle) is affected.
 
By making the transformation, 
\BEQ
\rho (t) = {\rm e}^{- i H t}\rho_{1}(t){\rm e}^{iHt},
\label{eliminate}
\EEQ
 the time dependence due to the Hamiltonian is eliminated; the 
quantity $\rho_{1}$ obeys the equation:
\BEQ
\frac{\partial \rho_{1}}{\partial t} + \frac{a}{T}\left(
f\left(\rho_{1}\right) - \frac{1}{N} {\rm Tr} f\left(\rho_{1}\right)
\right) = 0.
\label{rho1}
\EEQ

We notice that ``total disorder'',\IE
\BEQ
\rho_{1} = \frac{1}{N}
\label{chaos}
\EEQ
is a zero of $f - 1/N{\rm Tr}f$. Whether or not it is also an
attractor, depends on the functional form of $f$. (One can linearize
around \eq{chaos} in order to determine this; however, the
linearized version gives no information about the size of the basin of
attraction.) Due to the vanishing of $f - 1/N{\rm Tr}f$, the 
density matrix in equation~\eq{chaos} is stationary.

In order to make further progress, we now consider some simple
examples. The main technical simplification introduced is that we further
restrict the dimensionality of the state space: we take $N=2$; in this way, we
can take advantage of the properties of the algebra of quaternions 
(equivalently, of the Pauli matrices).

In two dimensions, a density matrix is of the form,
\BEQ
\rho = \frac{1}{2}\left( 1 + {\bf s}\cdot {\bf \sigma}\right),
\label{densitymatrix}
\EEQ
where ${\bf \sigma}$ stand for the Pauli matrices and ${\bf s}^{2}\leq 1$.
Clearly, since $\rho$ and $\rho_{1}$ are unitarily equivalent, $\rho_{1}$
can be written in the same form as \eq{densitymatrix}.
Consider now,
\begin{description}
\item[Example \# 1:]
\BEQ
f\left( \rho_{1}\right) = \rho_{1}^{2}.
\label{firstexample}
 \EEQ
Equation \eq{rho1} can be solved immediately. We have:
\BEQ
{\bf s}_{1}(t) = {\bf s}_{1}(0)\exp \left(\frac{-a t}{2 T}\right)
\label{rhosquared}
\EEQ
Clearly, only $a>0$ makes sense from the physical point of view,
since ${\bf s}_{1}^{2}\leq 1$. We take $a=2$ in order to normalize
the time dependence to $\exp -t/T$. (Evidently, such a choice of $a$
merely serves to normalize the time dependence in an appropriate
fashion; it does not influence the physics in any way.)
\item[Example \# 2:]
\BEQ
f\left(\rho_{1}\right) = \rho_{1}^{3}.
\label{secondexample}
\EEQ
Just like in the case of equation~\eq{firstexample}, the evolution
equation~\eq{rho1} can be solved in a closed form. The equation reads:
\BEQ
\frac{\partial {\bf s}_{1}}{\partial t} = - \frac{a}{8T}{\bf s}_{1}\left(
 3 +{\bf s}_{1}^{2}\right)
\label{rhocube}
\EEQ
Clearly, the direction of ${\bf s}_{1}$ is  constant and by taking the
scalar product of   equation~\eq{rhocube} with ${\bf s}_{1}$, one obtains
an equation for the magnitude of the
polarization,\footnote{``Polarization'' is used in a generalized sense.
In general, it is just a measure of the deviation from total disorder.} \VIZ
\BEQ
\frac{\partial {\bf s}_{1}^{2}}{\partial t} = \frac{- a}{4T}{\bf s}_{1}^{2}
\left( 3 + {\bf s}_{1}^{2} \right)
\label{magnitude}
\EEQ
The solution of eq.~\eq{magnitude} is:
\begin{eqnarray}   \label{msolved}
{\bf s}_{1}^{2}(t) & = & {\bf s}_{1}^{2}(0) {\rm e}^{-3at/4T} \nonumber \\
     & \times &\left[ 3 +{\bf s}_{1}^{2}(0)\left( 1 - {\rm e}^{ - 3at/4T} \right) \right]^{-1}
\end{eqnarray}
Clearly, only $a>0$ is physically acceptable and one can choose $a=4/3$
in order to standardize the time dependence.
Both examples considered so far are such that for physically acceptable
values of the parameter $a$, total disorder (equation~\eq{chaos}) is
an attractor and after times of the order of $T$ the quantum system will
find itself in the neighborhood of the attractor.
 
This need not be
the case, however. One can construct examples with other attractors.
It would be tempting to choose for $f$ something like
$\rho ( 1-\rho)$ in analogy with the logistics equation; 
presumably, such a term would drive the quantum
system towards a pure state. However,
that expression has no traceless part in two dimensions. One does not want
any dimensionality to be singled out, thus some more complicated
functional form has to be tried.
\item[{\em Example \# 3}:] Choose:
\BEQ
f(\rho) = \rho^{3} - \rho^{2}
\EEQ
Clearly, $f(\rho)=0$ for a pure state. Equation~\eq{rho1} can be solved in
a closed form for this case too. As before, it is sufficient to give
the time evolution of the magnitude of ${\bf s}$. One has:
\BEQ
{\bf s}^{2}(t)=\frac{{\bf s}^{2}(0)}{{\bf s}^{2}(0) +
\left( 1- {\bf s}^{2}(0)\right){\rm e}^{-at/4T}}.
\label{attractive}
\EEQ
In this case, both positive and negative values of $a$ lead to
physically acceptable results; however, the qualitative behavior
of the system depends crucially on the sign of $a$.
\begin{enumerate} 
\item If $a>0$, a pure state (${\bf s}^{2}=1$) is an attractor:
if initially ${\bf s}^{2} < 1$, the system will move towards 
a pure state unless initially ${\bf s}=0$. Any
point within the unit ball with the origin removed is within 
the basin of attraction of a pure state.
\item The situation is reversed for $a<0$: any point 
in the interior of the unit
ball (${\bf s}^{2}<1$) is in the basis of attraction of total disorder
(eq.~\eq{chaos}) and the surface is an unstable fixed point.
\end{enumerate}
This distinction is relevant from the point of view of experimental tests.
\end{description} 

The possibility of testing for the presence of terms proportional to $1/T$
in the evolution equation of a quantum system arises from the fact that
present and planned long baseline neutrino oscillation experiments 
take place on
length scales of the order of $d \simeq 10^{3}$km; hence, they should be
sensitive to characteristic times, $d = T \simeq 10^{-3}$s. This is
considerably larger than the time scales involved in typical
terrestrial experiments. (For comparison, a typical atomic
transition is characterized by times of the order of $1{\rm eV}^{-1}
\simeq 10^{-15}$s;
the time associated with the $K_{S}-K_{L}$ mass difference is
about $(\Delta m)^{-1}\simeq  2\times 10^{-10}$s.)

{\em In principle\/}, the test is a very simple one. We noticed that,
as a consequence of probability conservation, a  {\em``total disorder''\/}
characterized by eq.~\eq{chaos} is always a fixed point of 
equation~\eq{rho1}, and, hence, of the complete density operator.
From the examples considered, it is also likely that  total
disorder is an attractor unless matters are specially arranged,
as in Example \#3. Therefore, a likely test for the presence
of non linear terms in the evolution equation consists of a
search for spontaneous depolarization  as a function of time.
 
In order to make matters more quantitative, let us consider neutrino
oscillations with a non linear term  in the evolution equation
discussed in Example \#1 above. Neutrinos are particularly
advantageous from the point of view of testing for 
spontaneous decoherence, since their interaction with the
environment is generally negligibly small. Thus, one may be
able to distinguish between spontaneous and 
environmental~\cite{karolyhazy, zurek, omnes} decoherence.

Consider therefore a Hamiltonian of a two flavor system, say
$\left( \nu_{\mu}, \nu_{e} \right)$ \cite{kimpevsner},
with a Hamiltonian in the diagonal basis given as:
\BEQ
H= \frac{E_{1}+E_{2}}{2} + \frac{E_{1}-E_{2}}{2}\sigma_{3},
\label{hamiltonian}
\EEQ
and a mixing matrix,
\BEQ
U = {\rm e}^{i\theta \sigma_{2}}.
\label{mixing}
\EEQ
At momenta much higher than the rest masses of the neutrinos, the
energies are given by the expressions:
\BEQ
E_{i} \approx p + \frac{m_{i}^{2}}{2p} \qquad (i=1,2)
\EEQ
The density matrix in the flavor basis at production is given by:
\BEQ
\rho (0) = \frac{1}{2}\left( 1 + s_{3}(0) \sigma_{3}\right).
\label{initialrho}
\EEQ
The value of $s_{3}(0)$ in eq.~\eq{initialrho} equals $\pm 1$,
depending on whether $\nu_{\mu}$ or $\nu_{e}$ is produced.
Using the preceding equations and eq.~\eq{rhosquared}, one readily
obtains the density matrix at time $t$:
\begin{eqnarray} \label{density_matrix}
\rho(t) &=& 1/2 + 1/2 s_{3}(0){\rm e}^{-t/T}
\sigma_{3} \left( \cos^{2}\phi + \sin^{2} \phi \cos 4\theta \right)\nonumber \\
  &+& 1/2 s_{3}(0){\rm e}^{-t/T}\sin^{2} \phi \left(-\sigma_{1}+\sigma_{2}
\right), 
\end{eqnarray}
where
\[
\phi = \frac{\Delta m^{2}t}{4p}
\]
(In the limit of $T\rightarrow \infty $, eq.~\eq{density_matrix}, of
course, reproduces the standard result.) Similar expressions hold
whenever total disorder is an attractor, as in the second example
discussed above: in all such cases, the signature for a departure
from standard quantum mechanics is a damping of the
polarization.

Example \# 3 deserves special attention: the evolution
equation  has a fixed point at
${\bf s}^{2}=1$. It is generally assumed that weak interactions
produce neutrinos of a definite flavor, \IE in a pure state.
If the fixed point in Example \# 3 is a stable one, one has virtually no
chance of observing a deviation from standard quantum theory in 
a neutrino oscillation experiment, even though the evolution equation of
the density matrix contains non linear terms. Even if the fixed point
is unstable, one needs environmental perturbations in order to drive
the neutrino away from a pure state: the relevant Lyapunov
exponent may be too small  to make the presence of a
non linear term in the evolution equation observable.  The lesson
to be learnt from this example is that even though there may be
deviations from standard quantum theory present in the evolution
equation of the density operator, circumstances may conspire to
effectively hide that deviation from experimental scrutiny.

To summarize, long baseline neutrino oscillation experiments are likely to
provide an environment for testing the validity of standard
quantum theory, due to the unusually long distances involved in
such  experiments. From the examples considered here, it appears that,
if non linear terms are present in the evolution equation for the
density matrix, they are likely to lead to spontaneous decoherence.
Nevertheless, some caution is needed: there may be situations in
which the presence of deviations from standard quantum theory is
hidden from observation in certain experiments. One also has to
bear in mind that, in all the examples considered in this paper,
spontaneous decoherence leads to total disorder. This is not 
necessarily the general situation, however. One can think of
a number of non linear extensions of the evolution equation
for the density operator in which off diagonal elements go to
zero as time increases, but the density operator does not become
a multiple of the unit operator. 
 
{\em Acknowledgement.} This work was done during the authors' 
visit at the Dipartimento di Fisica, Universit\'{a} di Firenze.
We thank Roberto Casalbuoni, Director of the Department, for the
hospitality extended to us. We also thank Frigyes K\'{a}rolyh\'{a}zy
for several tutorial sessions on the problem of decoherence.
 
\end{document}